\documentstyle[12pt]{article}

\catcode`\@=11
\long\def\@makefntext#1{
\protect\noindent \hbox to 3.2pt {\hskip-.9pt
$^{{\ninerm\@thefnmark}}$\hfil}#1\hfill}		

\def\@makefnmark{\hbox to 0pt{$^{\@thefnmark}$\hss}}  

\def\ps@myheadings{\let\@mkboth\@gobbletwo
\def\@oddhead{\hbox{}
\rightmark\hfil\ninerm\thepage}
\def\@oddfoot{}\def\@evenhead{\ninerm\thepage\hfil
\leftmark\hbox{}}\def\@evenfoot{}
\def\sectionmark##1{}\def\subsectionmark##1{}}

\renewcommand{\thefootnote}{\fnsymbol{footnote}}

\newcounter{sectionc}\newcounter{subsectionc}\newcounter{subsubsectionc}
\renewcommand{\section}[1] {\vspace*{0.6cm}\addtocounter{sectionc}{1}
\setcounter{subsectionc}{0}\setcounter{subsubsectionc}{0}\noindent
	{\normalsize\bf\thesectionc. #1}\par\vspace*{0.4cm}}
\renewcommand{\subsection}[1] {\vspace*{0.6cm}\addtocounter{subsectionc}{1}
	\setcounter{subsubsectionc}{0}\noindent
	{\normalsize\it\thesectionc.\thesubsectionc. #1}\par\vspace*{0.4cm}}
\renewcommand{\subsubsection}[1]
{\vspace*{0.6cm}\addtocounter{subsubsectionc}{1}
	\noindent {\normalsize\rm\thesectionc.\thesubsectionc.\thesubsubsectionc.
	#1}\par\vspace*{0.4cm}}

\newcounter{appendixc}
\newcounter{subappendixc}[appendixc]
\newcounter{subsubappendixc}[subappendixc]

\renewcommand{\appendix}[1] {\vspace*{0.6cm}
        \refstepcounter{appendixc}
        \setcounter{figure}{0}
        \setcounter{table}{0}
        \setcounter{equation}{0}
        \renewcommand{\thefigure}{\Alph{appendixc}.\arabic{figure}}
        \renewcommand{\thetable}{\Alph{appendixc}.\arabic{table}}
        \renewcommand{\theappendixc}{\Alph{appendixc}}
        \renewcommand{\theequation}{\Alph{appendixc}.\arabic{equation}}
        \noindent{\bf Appendix \theappendixc #1}\par\vspace*{0.4cm}}

\def\abstracts#1{{

\centering{\begin{minipage}{12.2truecm}\footnotesize\baselineskip=12pt\noindent
	\centerline{\footnotesize ABSTRACT}\vspace*{0.3cm}
	\parindent=0pt #1
	\end{minipage}}\par}}

\renewenvironment{thebibliography}[1]
	{\begin{list}{\arabic{enumi}.}
	{\usecounter{enumi}\setlength{\parsep}{0pt}
\setlength{\leftmargin 1.25cm}{\rightmargin 0pt}
	 \setlength{\itemsep}{0pt} \settowidth
	{\labelwidth}{#1.}\sloppy}}{\end{list}}

\newcounter{itemlistc}
\newcounter{romanlistc}
\newcounter{alphlistc}
\newcounter{arabiclistc}

\newcommand{\fcaption}[1]{
        \refstepcounter{figure}
        \setbox\@tempboxa = \hbox{\footnotesize Fig.~\thefigure. #1}
        \ifdim \wd\@tempboxa > 6in
           {\begin{center}
        \parbox{6in}{\footnotesize\baselineskip=12pt Fig.~\thefigure. #1}
            \end{center}}
        \else
             {\begin{center}
             {\footnotesize Fig.~\thefigure. #1}
              \end{center}}
        \fi}

\newcommand{\tcaption}[1]{
        \refstepcounter{table}
        \setbox\@tempboxa = \hbox{\footnotesize Table~\thetable. #1}
        \ifdim \wd\@tempboxa > 6in
           {\begin{center}
        \parbox{6in}{\footnotesize\baselineskip=12pt Table~\thetable. #1}
            \end{center}}
        \else
             {\begin{center}
             {\footnotesize Table~\thetable. #1}
              \end{center}}
        \fi}

\def\@citex[#1]#2{\if@filesw\immediate\write\@auxout
	{\string\citation{#2}}\fi
\def\@citea{}\@cite{\@for\@citeb:=#2\do
	{\@citea\def\@citea{,}\@ifundefined
	{b@\@citeb}{{\bf ?}\@warning
	{Citation `\@citeb' on page \thepage \space undefined}}
	{\csname b@\@citeb\endcsname}}}{#1}}

\newif\if@cghi
\def\cite{\@cghitrue\@ifnextchar [{\@tempswatrue
	\@citex}{\@tempswafalse\@citex[]}}
\def\citelow{\@cghifalse\@ifnextchar [{\@tempswatrue
	\@citex}{\@tempswafalse\@citex[]}}
\def\@cite#1#2{{$\null^{#1}$\if@tempswa\typeout
	{IJCGA warning: optional citation argument
	ignored: `#2'} \fi}}

 1
 1
 1

\font\ninerm=cmr9

\begin{document}

\centerline{\normalsize\bf WILSON LOOP APPROACH}
\baselineskip=22pt
\centerline{\normalsize\bf TO THE $q \bar{q}$ INTERACTION PROBLEM}
\centerline{\footnotesize N. BRAMBILLA
 and G. M. PROSPERI\footnote{Presenting author}}
\baselineskip=13pt
\centerline{\footnotesize\it Dipartimento di Fisica dell'Universit\`a, Milano}
\baselineskip=12pt
\centerline{\footnotesize\it INFN, Sezione di Milano, Via Celoria 16, 20133
 Milano}
\centerline{\footnotesize E-mail: prosperi@mi.infn.it}

\vspace*{0.9cm}
\abstracts{ It is shown that the semirelativistic $q \bar{q}$ potential,
 the relativistic flux tube model and a confining Bethe--Salpeter
 equation can be derived from QCD first principles in a unified
point of view.}

\normalsize\baselineskip=15pt
\setcounter{footnote}{0}
\renewcommand{\thefootnote}{\alph{footnote}}
\vskip 1truecm
In this paper we want to show how, starting from  the same standard
  evaluation of the Wilson loop integral and using
 similar techniques, it is possible to justify three different
  approaches to a treatment  of the $q \bar{q}$ interaction
 on the basis of QCD alone,  without
making
 any ad hoc phenomenological
 hypothesis.\par
The three approaches are the derivation of a semirelativistic
potential,  the relativistic flux tube model and the
Bethe--Salpeter equation. For simplicity, having  here mainly a
methodological  purpose,  we shall neglect spin in the last two cases.
 The modifications arising by the  consideration of the quark
spin  shall be discussed in the paper presented by N.
Brambilla~\cite{bramb}
 to which in a sense this one serves as an introduction.\par
As usual the Wilson loop integral is defined by
\begin{equation}
W = {1\over 3}
 \langle {\rm Tr} {\rm P}_{\Gamma}
 \exp i g \{ \oint_{\Gamma} dx^{\mu} A_{\mu} \}
        \rangle  \
\label{eq:loop}
  \end{equation}
where  the loop $\Gamma$ is supposed
 made by a quark world line ($ \Gamma _1 $), an
antiquark world line ($ \Gamma _2 $)  and  two
straight lines connecting the initial and the final points of the two world
lines ($y_1, y_2$ and $x_1 , x_2$).
The basic  assumption is
 \begin{equation}
i \ln W = i (\ln W)_{\rm pert} + \sigma S_{\rm min}\,  ,
\label{eq:iniz}
   \end{equation}
$(\ln W)_{\rm pert}$ being the perturbative contribution to $\ln W$ and
$S_{\rm min}$ the minimum area enclosed by $\Gamma$.
At lowest order in $g^2$  one has
\begin{eqnarray}
i (\ln W )_{\rm pert}& &  = {4\over 3} g^2 \int_{y_{10}}^{x_{10}}
d t_1 \int_{y_{20}}^{x_{20}} d t_2 { d z_1^\mu \over dt_1}
 {d z_2^\nu \over d t _2} D_{\mu \nu}(z_1-z_2) \nonumber\\
& & + {2\over 3} g^2
\int_{y_{10}}^{x_{10}} d t_1 \int_{y_{10}}^{x_{10}} d t_1^\prime
 { d z_1 \over d t_1} { d z_1^\prime \over d t_1^\prime}
D_{\mu \nu} (z_1- z_1^\prime) +\nonumber \\
& &+{2\over 3} g^2 \int_{y_{20}}^{x_{20}} d t_2 \int_{y_{20}}^{x_{20}}
 d t_2^\prime { d z_2^\mu \over d t_2} { d z_2^{\prime\nu}\over d
t_2^\prime } D_{\mu \nu} (z_2-z_2^\prime) =\nonumber\\
& &=\int_{t_i}^{t_f} \big \{ - {4\over 3} {\alpha_s \over r} [1 -
{1\over 2} (\delta^{hk} + \hat{r}^h \hat{r}^k ) v_1^h v_2^k +\dots
\big \},
\end{eqnarray}
$D_{\mu \nu}$  being the gluon propagator; $z_1= z_1(t_1)$ and
$z_2=z_2(t_2)$ the quark and the antiquark worldlines
 ( with $z_{j0}= t_j$, ${\bf z}_j= {\bf z}_j (t_j) $, $ z_{j0}^\prime=
t_j^\prime $, ${\bf z}_j^\prime ={\bf z}_j(t_j^\prime)$ and
 ${\bf v}_1$ and $ {\bf v}_2 $ the corresponding velocities;
 $ {\bf r}= {\bf z}_1- {\bf z}_2$ the relative position).
 To make the second step in (3) one has to neglect the self--energy
terms; to set $y_1^0 =y_2^0= t_i$, $x_1^0=x_2^0= t_f $,
$ t_1= t-{\tau \over 2} $, $t_2= t -{\tau \over 2}$;
 to expand $z_1$ and $ z_2$  in the relative time $\tau$
 and to integrate over this.
 Furthermore, we use
for $S_{\rm min}$ the straight line approximation, consisting in
replacing
$S_{\rm min}$ with the surface spanned by the straight lines connecting
equal time points on the quark and the antiquark worldlines. In
 practice we
write
   \begin{eqnarray}
       S_{\rm min}  & &  \cong   \int_{t_{\rm i}}^{t_{\rm f}}
 dt\,  r \int_0^1 ds [1-(s { d {\bf z}_{1 {\rm T}} \over d t}
           + (1-s) {d {\bf z}_{2 {\rm T}} \over d t})^2 ]^{1 \over 2}=
\nonumber \\
& & = \int_{t_i}^{t_f} dt [ 1 -{1\over 6} ( {\bf v}_{1 {\rm T}}^2 +
{\bf v}_{2 {\rm T}}^2 + {\bf v}_{1{\rm T}} \cdot {\bf v}_{2{\rm T}} )
 + \dots ]
\label{eq:minform1}
        \end{eqnarray}
$ { d {\bf z}_{ j {\rm T}} \over d t }$ and $ {\bf v}_{ j {\rm T}}$
 denoting the transverse velocities :
 $ v_{j {\rm T}}^h = (\delta^{hk}- \hat{r}^h \hat{r}^k ) v_j^k$.\par
 The basic object  that we consider in our discussion is the
usual gauge invariant quark--antiquark propagator
\begin{eqnarray}
G^{gi}_4(x_1,x_2,y_1,y_2) &=&
\frac{1}{3}\langle0|{\rm T}\psi_2^c(x_2)U(x_2,x_1)\psi_1(x_1)
\overline{\psi}_1(y_1)U(y_1,y_2)  \overline{\psi}_2^c(y_2)
|0\rangle =
\nonumber\\
&=& \frac{1}{3} {\rm Tr} \langle U(x_2,x_1)
 S_1(x_1,y_1;A) U(y_1,y_2) C^{-1}
S_2(y_2,x_2;A) C \rangle
\label{eq:gauginv}
\end{eqnarray}
where  $C$ denotes the charge-conjugation
matrix, $U$
the path-ordered gauge string
\begin{equation}
U(b,a)= {\rm P}_{ba}  \exp  \left(ig\int_a^b dx^{\mu} \, A_{\mu}(x) \right)
\label{eq:col}
\end{equation}
(the integration path being the straight line joining $a$ to $b$),
 $S_1$ and $S_2$ the single quark propagators in the
external gauge field $A^{\mu}$ which are supposed
 to be defined
 by  the equation
\begin{equation}
[ i\gamma^\mu (\partial_\mu - i g A_\mu)  -m] S(x,y;A) =\delta^4(x-y)
\label{eq:propdir}
\end{equation}
and the appropriate  boundary conditions.\par
The various mentioned approaches to the treatment of the $ q\bar{q}$
system correspond to different manipulations of (7) and (5).
It is common to all cases the explicit resolution of (7)
in terms of a  path integral.

\section{Semirelativistic potential ~\cite{luch,bak}}
For $x^0 > y^0$ by performing on (7) a
Foldy--Wouthuysen transformation one can replace the $ 4\times 4$
 Dirac type propagator $S(x,y;A)$ by a $ 2 \times 2$ Pauli type
propagator $ K(x,y;A)$ which satisfies a  Schr\"odinger  type equation
 with a  hamiltonian expressed as a $ {1\over m^2}$ expansion. Solving
this last equation by a path integration technique and using the  expression
 so obtained in (5) (again for $x_1^0= x_2^0=t_f$,
 $y_1^0=y_2^0=t_i$  with
 $t_f-t_i > 0 $ and large) one arrives eventually to the two--particle
 Pauli type propagator
\begin{eqnarray}
& & K({\bf x}_1, {\bf x}_2, {\bf y}_1, {\bf y}_2; t_{\rm f} - t_{\rm i})=
\int_{{\bf z}_1(t_{\rm i})={\bf y}_1}^{{\bf z}_1(t_{\rm f})={\bf x}_1}
    {\cal D}{\bf z}_1 {\cal D} {\bf p}_1
\int_{{\bf z}_2(t_{\rm i})={\bf y}_2}^{{\bf z}_2(t_{\rm f})={\bf x}_2}
    {\cal D}{\bf z}_2 {\cal D} {\bf p}_2 \nonumber \\
& &  \exp \{i\int_{t_{\rm i}}^{t_{\rm f}} dt\,
 \sum_{j=1}^2
[{\bf p}_j \cdot \dot{{\bf z}}_j -m_j-
\frac{{\bf p}^2_j}{2m_j}+\frac{{\bf p}^4_j}{8m_j^3} ]\}
\langle \frac{1}{3}
{\rm Tr \, T_s \, P} \exp \{ig\oint_{\Gamma} dx^{\mu} \,
A_{\mu}(x) \nonumber \\
& & +\sum_{j=1}^2\frac{ig}{m_j} \int_{{\Gamma}_j} dx^{\mu}
  (S_j^l \hat{F}_{l{\mu}}(x) -\frac{1}{2m_j}
S_j^l\varepsilon^{lkr}p_j^k F_{{\mu}r}(x)-
\frac{1}{8m_j} D^{\nu}F_{{\nu}{\mu}}(x)
 ) \}  \rangle \> .
\label{eq:path1}
\end{eqnarray}
where ${\rm T_s}$ is the time-ordering prescription for the
 spin matrices alone.
The semirelativistic potential is obtained  by comparing (8)
 with the path integral solution of the two particle Schr\"odinger
 equation, having  used the second step of (3) and (4) and having
reduced  the spin dependent terms to functional derivatives
 of $ \ln W$. The final result is
\begin{eqnarray}
V&  = & -  \frac{4}{3}
 \frac{{\alpha}_s}{r} + \sigma r -{4\over 3} {\alpha_{\rm s}^2 \over 4 \pi}
 {1\over r} [{66- 4 N_{f} \over 3}
 (\ln \mu r + \gamma) +A]+\nonumber \\
&  + &{1\over m_1 m_2} \{ {4\over 3}
 {\alpha_s \over r} ( \delta^{hk}+ \hat{r}^h \hat{r}^k)
 p_1^h p_2^k \}_{\rm Weyl\, ord}   \\
& -& \sum_{j=1}^2 {1\over 6 m_j^2} \{ \sigma r {\bf p}_{j {\rm T}}
\}_{\rm Weyl\, ord} -{1\over 6 m_1 m_2} \{ \sigma r {\bf p}_{1 {\rm T}}
 \cdot {\bf p}_{2 {\rm T}} \}_{\rm Weyl\, ord}+\nonumber \\
& & +\frac{1}{8} \sum_{j=1,2} {1\over m_j^2}
\nabla^2 \left( - \frac{4}{3} \frac{\alpha_s}{r} + \sigma r \right)
+
   \left(
 \frac{4}{6} \frac{\alpha_s}{r^3} -
\frac{\sigma}{2 r} \right) \sum_{j=1,2} \frac{1}{m_j^2} {\bf S}_j \cdot
{\bf L}_j
 + \frac{1}{m_1m_2} \frac{4}{3} \frac{\alpha_s}{r^3} \times\nonumber \\
& & ({\bf S}_2
\cdot{\bf L}_1  + {\bf S}_1 \cdot{\bf L}_2 ) +
 {4 \alpha_{\rm s}\over 3 m_1 m_2} \left [
 (\frac{3}{r^5} ({\bf S}_1 \cdot {\bf r})({\bf S}_2 \cdot
{\bf r}) - {{\bf S}_1 \cdot {\bf S}_2\over r^3}) +
\frac{8\pi}{3} \delta^3({\bf r}) {\bf S}_1 \cdot {\bf S}_2 \right ]
\nonumber
\label{eq:trenb}
\end{eqnarray}

\section{Relativistic flux tube model ~\cite{olss}}

Let us neglect in Eq.(8) the spin--dependent terms
 and replace the ${1\over m^2}$ expansion of the kinetic term
by its exact relativistic
 expression
\begin{eqnarray}
& &K({\bf x}_1, {\bf x}_2;  {\bf y}_1, {\bf y}_2; t_{\rm f} - t_{\rm i}) =
\nonumber \\
& &\int {\cal D}{\bf z}_1 {\cal D}{\bf p}_1
\int {\cal D}{\bf z}_2 {\cal D}{\bf p}_2
\exp \left\{i \left[ \int_{t_{\rm i}}^{t_{\rm f}} dt \sum_{j=1}^2 ({\bf p}_j
\cdot {\dot{\bf z}_j} - \sqrt{m_j^2 + {\bf p}_j^2})
\right] + \ln W \right\} .
\label{eq:alfa}
\end{eqnarray}
Let us also neglect for simplicity
 $ i (\ln W)_{\rm pert}$  in (2)  and assume that a sensible
 approximation is obtained even in the  relativistic case postulating
  the first line  of (4) in the center--of--mass frame
 of the two particles. Then, integrating on the momenta one obtains
 the ordinary  lagrangian
\begin{equation}
 L   = - \sum_{j=1}^2 m_j \sqrt{1- \dot {\bf z}_j^2 }
 - \sigma
r \int_0^1 ds [1 - ( s \dot{\bf z}_{1{\rm T}} + (1-s) \dot{{\bf
z}}_{2{\rm T}} )^2 ]^{1/2}.
\label{eq:beta}
\end{equation}
This coincides with the relativistic flux--tube model
lagrangian~\cite{olss}.\par
 From (12) is not possible to obtain even a classical
 hamiltonian in a closed form, due to the complicate velocity dependence.
However, in terms of an expansion in ${\sigma \over m^2}$ we have  (we
 assume $m_1=m_2=m$ for simplicity)
\begin{equation}
{\cal{H}}({\bf r}, {\bf q}) = 2 \sqrt{m^2+q^2} +{\sigma r\over 2}
 \Big [{\sqrt{m^2+q^2}\over q_{\rm T}}\, {\rm arcsin}{q_{\rm T}
 \over \sqrt{m^2+q^2}}
+\sqrt{{m^2+ q_{\rm r}^2\over m^2+q^2}}\Big ]+\dots
\end{equation}
with ${\bf r}= {\bf z}_{1{\rm CM}}- {\bf z}_{2{\rm CM}}$,
${\bf q}={\bf p}_{1 {\rm CM}}=-{\bf p}_{2 {\rm CM}}$, ${\bf q}_{\rm r}=
(\hat{\bf r}\cdot {\bf q})/ {\hat{\bf r}}$ and  $q^h_{\rm T}= (\delta^{hk}
 -\hat{r}^h \hat{r}^k ) q^k$.
{}From this a quantum hamiltonian can  be immediately obtained
 by setting
\begin{equation}
 \langle  {\bf k}' \vert H_{\rm FT} \vert {\bf k}\rangle=
 \int {{\rm d}{\bf r} \over (2 \pi)^3} e^{i ({\bf k}-{\bf k}')
\cdot {\bf r}} \,\,
 {\cal {H}}({\bf r}, {{\bf k}'+
{\bf k}\over 2}),
\label{eq:delta}
\end{equation}
Then by an expansion in ${1\over m^2}$ a semirelativistic
 hamiltonian  can be obviously recovered with a confining
 potential
 given by the spin--independent part of (9).

\section{ Bethe--Salpeter equation ~\cite{bs,tion}}

Let us go back  to the
quantity analogous to (5)
 for spinless quarks
 and  in it use  the covariant representation for the quark propagator
 in an external gauge field
\begin{equation}
\Delta (x,y\vert A)={- i\over 2} \int_0^{\infty} d \tau
 \int_{z(0)=y}^{z(\tau)=x} {\cal D}z {\rm P}
{\rm exp}\,  i\int_0^{\tau} {\rm d} \tau^\prime \{ -{1\over 2}
 [ ({d z \over d \tau^\prime})^2 +m^2] - g z^{\mu \prime} A_{\mu} (z) \}
\label{eq:prop}
\end{equation}

 In place of (8)
 we find
\begin{eqnarray}
& &
G_4(x_1,x_2; y_1, y_2) =({-i\over 2})^2 \int_0^{\infty} d\tau_1
\int_0^{\infty} d\tau_2 \int_{z_1(0)=y}^{z_1(t_1)=x_1} {\cal
 D}z_1 \int_{z_2(0)=y_2}^{z_2(\tau_2)=x_2} {\cal D} z_2
\nonumber \\
& & {\rm exp}\big \{
 {-i \over 2}  \int_0^{\tau_1} d \tau_1' [
 ({d z_1\over d\tau_1'})^2 + m_1^2 ] -{i\over 2} \int_0^{\tau_2} d \tau_2'
 [ ({d z_2\over  d\tau_2'})^2 +m_2^2 ] +\ln W
\big \}
\label{eq:g4bs}
\end{eqnarray}
where  the  path
 connecting $y$ with $x$ is now written as
$ z^{\mu} = z^{\mu}(\tau) $,
 in terms of an independent parameter
 $\tau$ (rather than   the time $t$)
 and $z'$ stands for $z(\tau')$.
Then assuming again the first steps of (3) and (4)  in the center of
mass frame (after rewriting in terms of $\tau_1$ and $\tau_2$),
 replacing them in (5) and performing appropriate manipulations,
 one can obtain an inhomogeneous Bethe--Salpeter equation
 with a kernel resulting by the sum $ I= I_{\rm pert}+
I_{\rm conf}$ of a perturbative part  and a confinement one. This last
  in the momentum representation can be written as
\begin{equation}
\hat{I}_{\rm conf} (p_1', p_2'; p_1,p_2)  =
 {1\over (2 \pi)^3} \int d^3 {\bf r} e^{i ({\bf k}'- {\bf k})\cdot
 {\bf r}}  J({\bf r}, { { p}^{'}_1+p_1 \over 2},
 { p_2'+p_2 \over 2})
\label{eq:bsform}
\end{equation}
($p_1^\prime +p_2^\prime= p_1 +p_2 ,\,  {\bf p}_1= - {\bf p}_2= {\bf k}$,
 $ {\bf p}_1^\prime= - {\bf p}_2^\prime={\bf k}^\prime $) with
\begin{eqnarray}
J({\bf r}, q_1, q_2)& = & (2 \pi)^3 {\sigma r \over 2}
 {1\over q_{10} +q_{20}} [ q_{20}^2 \sqrt{ q_{10}^2 -{\bf q}_{\rm T}^2}
+ q_{10}^2  \sqrt{q_{20}^2 -{\bf q}^2_{\rm T}} +\nonumber \\
& +& { q_{10}^2 q_{20}^2 \over \vert {\bf q}_{\rm T}\vert }
 ( {\rm arcsin }{ \vert {\bf q}_{\rm T} \vert \over
 \vert q_{10}\vert } + {\rm arcsin } { \vert {\bf q}_{\rm T}\vert
 \over \vert q_{20}\vert } )] + O({\sigma^2 \over m^4})
\label{eq:kerndef}
\end{eqnarray}
Essential steps in the derivation are the equation
\begin{equation}
S_{\rm min} =\int_0^{s_1} d \tau_1 \int_0^{s_2}
 d \tau_2 \delta(z_{10} - z_{20} ) \vert {\bf z}_1 -{\bf z}_2 \vert
 \int_0^1 d s  \{ \dot{z}_{10}^2 \dot{z}_{20}^2 - ( s {\bf z}_{1 {\rm
T}} \dot{z}_{20} + ( 1-s) \dot{\bf z}_{2
{\rm T}} \dot{z}_{10} )^2 \}^{1\over 2}
\end{equation}
equivalent to (3) and the recurrence identity
\begin{equation}
\exp i \int_0^{s_1} d \tau_1 \int_0^{s_2} d \tau_2 f(z_1,z_2)=
1 + i\int_0^{s_1} d \tau_1 \int_0^{s_2} d \tau_2 f(z_1, z_2)
\exp i \int_0^{\tau_1} d \tau_1^\prime \int_0^{s_2} d \tau_2^\prime
 f(z_1^\prime, z_2^\prime).
\end{equation}
Notice that, according to a standard procedure, the BS kernel
 $\hat{I}$  can be associated  with a relativistic potential
 (to be used in the Salpeter equation) given by
\begin{equation}
\langle {\bf k}' \vert V \vert {\bf k} \rangle ={1\over
 (2 \pi)^3} { m_1 m_2 \over \sqrt{ w_1({\bf k}) w_2({\bf k})
 w_1({\bf k}') w_2({\bf k}')}} \hat{I}_{\rm inst} ({\bf k}', {\bf k})
\end{equation}
where $w_j({\bf k})= \sqrt{m_j^2+{\bf k}^2}$
 and the instantaneous kernel $ \hat{I}_{\rm inst}$ is obtained
 from $\hat{I}$ by setting
$p_{j0} = p_{jo}^\prime = {1\over 2} ( w_j({\bf k}) + w_j({\bf
k}^\prime )$. Obviously the resulting hamiltonian coincides  with
(12), (13).

\end{document}